\documentstyle[prl,aps]{revtex}
\newcommand{\be}{\begin{equation}}
\newcommand{\ee}{\end{equation}}
\newcommand{\ba}{\begin{eqnarray}}
\newcommand{\ea}{\end{eqnarray}}
\begin{document}
\draft
\twocolumn
\title{On the trace identity in a model with broken symmetry}
\author{Daniel H.T. 
Franco\thanks{
Supported by the {\it Conselho Nacional de Desenvolvimento 
Cient{\'\i}fico e
Tecnol{\'o}gico (CNPq)}
under the contract number 201142/97-0. 
E-mail: franco@manon.he.sissa.it}}
\address{{\normalsize {\it Scuola Internazionale Superiore
di Studi Avanzati (SISSA),\\
Via Beirut 2-4 - 34014 - Trieste - Italy}}}
\date{\today}
\maketitle
\begin{abstract}
Considering the simple chiral fermion meson model
when the chiral symmetry is explicitly broken, we show the
validity of a trace identity -- to all orders of perturbation
theory -- playing the role of a Callan-Symanzik equation 
and which allows us to identify directly
the breaking of dilatations with the trace of the
energy-momentum tensor.
More precisely, by coupling
the quantum field theory considered to a classical curved space
background, represented by the non-propagating external vielbein field,
we can express the conservation of the energy-momentum tensor through
the Ward identity which characterizes the invariance of the theory under
the diffeomorphisms. Our ``Callan-Symanzik equation'' then is the
anomalous Ward identity for the trace of the energy-momentum tensor,
the so-called ``trace identity''.
\end{abstract}

\bigskip


\section{Introduction}


A classical field theory is scale invariant precisely when it has
the following properties: it contains no mass terms and all coupling
constants are dimensionless. For a theory of this type it is possible
to define an energy-momentum tensor with vanishing trace
(see~\cite{wcz} for a review). On the other hand, it is natural
to expect that the scale symmetry be
broken to provide us a scale to live on. There are many different
ways to break scale invariance. The most ordinary way is that
scale invariance is broken due to quantum effects.

As it is well known, in the context of a renormalized
perturbation theory,
integrals associated with the Feynman graphs are generally UV
divergents. To give a proper meaning to such expressions we have 
to adopt a suitable
subtraction scheme. The effect of the latter is to render the integrals
convergent. The renormalization is performed through the addition of  
a certain number of counterterms to the initial action considered, whose
coefficients are left arbitrary. These coefficients have however to be
fixed by a set of normalization conditions, which are applied at a 
certain momentum scale. In this point, the important feature is the
appearance of a normalization parameter $\kappa$; it has the dimension
of a mass and will thus cause the breaking of scale invariance. The 
natural choice $\kappa=m$, where $m$ is the physical mass, cannot
be used because it leads to singularities at the massless limit. 
In this way, there 
is no way to preserve quantum scale invariance at the massless limit:
scale invariance is anomalous\footnote{Breaking of quatum scale
invariance by an anomaly is quite general.
Only some models remain scale
invariant at the quantum level (see for 
instance~\cite{lps,bc,ms,lps1,odhp}). 
Within perturbative theory, a necessary
and sufficient
condition to obtain the ``scale invariance'',
it is the vanishing to all orders of the $\beta$-functions
-- anomalous dimensions are allowed to be different of zero. The
latter corresponding to field redefinitions, are physically
trivial and hence vanish on the 
mass-shell~\cite{piguet}.}\cite{ccj,cs,sw,fmw,cdj}.
This is called the trace anomaly induced by radiative
corrections.

In order to get insight into this problem, a systematic study of
renormalization properties can be achieved via the Callan-Symanzik
(CS) equation~\cite{cs}. It describes the behaviour of the quantum
theory under
scale transformations, being successfully applied in
symmetric models. 
If the theory is broken
the construction of the CS equation is 
more involved~\cite{becchi,kraus1,mpw}.
Due to the shifts by constant amounts in certain fields,
the dilatation Ward operator does not will commute with the Ward operator
for broken symmetry. Therefore the breaking of dilatations is not
symmetric but has a certain covariance under the symmetry transformations
already at the classical level. As alluded in Ref.~\cite{kraus1}, 
to identify the breaking of
dilatations with the trace of the energy-momentum tensor is thus
complicated at the quantum level. In addition,
in a purely physical parametrization, the effect of the
breaking induces the appearance of physical mass $\beta$-functions
~\cite{becchi,kraus1,mpw}, a situation which asks for a deeper
understanding. Of a particular kind is the case for a realistic
supersymmetric gauge field theory~\cite{mpw}. 

The purpose of the present paper is to supplement the works of
\cite{becchi,kraus1,mpw} and, exploiting the techniques developed
in~\cite{odhp}, to provide an alternative way of deriving the
CS equation,
which allows us to identify the breaking
of dilatations
with the trace of the energy-momentum tensor
in a model with broken symmetry. More precisely, by coupling
the quantum field theory considered to a classical curved space
background, represented by the non-propagating external vielbein field,
we can express the conservation of the energy-momentum tensor through
the Ward identity which characterizes the invariance of the theory under
the diffeomorphisms. Our ``Callan-Symanzik equation'' then is the
anomalous Ward identity for the trace of the energy-momentum tensor,
the so-called ``trace identity''~\footnote{There exist studies 
in the literature concerning the local
dilatation properties -- general coordinate transformations
in the Weyl's sheme -- as discussed by G. Bandelloni et 
al~\cite{bbbc}, where a more difficult task was
taken up: the radiative mass generation due the 
trace anomaly. Even though their approach is related to ours,
here, instead of introducing an external dilatation field
beyond the external metric or vielbein field, we only
consider the latter.}. As a by-product, the approach will
allow us to bring to an end that the $\beta$ and $\gamma$-functions 
in curved space-time are the same as in flat one.

Since we are working with an external vielbein which is not necessary flat,
our results hold for a curved manifold, as long as its topology remains that
of flat ${\cal R}^4$, with asymptotically vanishing curvature. It is the
latter two restrictions which allow us to use the general results of
renormalization theory, established in flat space. Indeed, we may then
expand in the powers of $\bar{e}_\mu^m$ = $e_\mu ^m$ $-$ $\delta_\mu^m$,
considering $\bar{e}_\mu^m$ as a classical background field in flat 
${\cal R}^4$, and thus make use of the 
general theorems of renormalization theory
actually proved for flat space-time \cite{qap,Zimmerman}. 

We shall consider the model of chiral fermion meson with explicit
breaking of the chiral symmetry. In contrast with the Ref.~\cite{bert},
which have considered the explicit breaking term in order to be
able to treat the massless Goldstone particle in its massless limit,
we are interested in the case where the pion fields are massive. The
importance of this lies in fact that the chiral fermion meson model
allows for
the possibility of a simultaneous description of the baryonic and
mesonic low energy sector in hadron physics\footnote{See {\it e.g.}
\cite{hdafc} for current study of the model in a somewhat
phenomenological basis when temperature effect are considered, 
in absence of fermions.}. Consistency conditions for chiral symmetry
and scale invariance, in absence of fermions and charged pions, were
studied in~\cite{jackiw}\footnote{I thank Prof. R. Jackiw for drawing
these works to my attention.}.

The outline of the paper is as follow. The model in a curved Riemannian
manifold described in terms of external vielbein and spin connection
fields, is introduced in Section 2, together with its symmetries. The trace
identity for the classical theory is derived in Section 3. The extension of
this identity to the quantum level is discussed in Section 4, followed by
a summary. For the sake of completeness, we add
two appendices: The renormalizability of the model is sketched in the
appendix A. Assuming the limit of
flat space-time, the CS equation is derived
in the appendix B.


\section{Generalities of the model}

\label{sect2}

In this Section, we give a brief description of the model in a
curved space-time. 
The chiral fermion meson model involves a fermionic isodoublet,
$\psi$, of zero mass, a scalar sigma field, $\sigma$, and a triplet
of charged pseudscalar pion fields, $\pi^a$.
Because fermions are represented by spinor
fields which are subjected to the Lorentz group -- and not to the
diffeomorphisms group -- we must refer these fields to the tangent
frame and treat the fermions as scalars with respect to the
diffeomorphisms. We achieve this with the help of the vierbein.
Space-time is a 4-dimensional Riemannian manifold ${\cal M}$, with
coordinates $x^\mu$, $\mu =0,1,2,3$. It is described by a vierbein field 
$e_\mu ^m(x)$ and its inverse $e_m^\mu (x)$, $\mu$ being a world index and 
$m$ a tangent space index. The spin connection $\omega _\mu^{mn}(x)$ is
not an independent field, but depend on the vierbein due to the vanishing
torsion condition: $\omega_{lmn}=\frac 12
\left(\Omega_{lmn}+\Omega_{mnl}-\Omega_{nlm}\right)$,
with $\Omega_{lmn}=e_l^\mu e_m^\nu(\partial_\mu e_{\nu n}-
\partial_\nu e_{\mu n})$.

The metric tensor reads $g_{\mu \nu }\left( x\right) 
=\eta _{mn}\,e_\mu ^m\left( x\right) e_\nu
^n\left( x\right)$; 
$\eta _{mn}$ being the tangent space flat metric. 
We denote by \thinspace$e$ \thinspace the 
determinant of $e_\mu ^m$. As explained in the 
Introduction, we assume the manifold ${\cal M}$ to be
topologically equivalent to ${\cal R}^4$ and asymptotically flat.

We shall take into account the physically interesting case 
when the chiral
symmetry is broken adding a linear term to the
action. The linear breaking term implies that the quantum 
$\sigma$ field has a nonvanishing vacuum expectation value $\langle 
\sigma
\rangle =v\,$. If one whishes to interpret the theory in terms
of particles, it is necessary to perform a field translation
$\sigma \rightarrow \sigma+v$ such that 
$\left. \frac {\delta \Sigma} {\delta \sigma}
\right|_{\sigma=v}=0$.
As an effect the mass degeneracy between fields $\sigma$ and
$\pi^a$ disappears and the fermionic isodoublet $\psi$ acquires a mass.

The corresponding action in a curved manifold is given by:
{\small{
\begin{eqnarray}
\Sigma &=&\int d^4x\,e\,\left\{\bar{\psi}\,i\gamma ^\mu {\cal D}_\mu \psi
+\frac 12 \left(\partial _\mu \sigma \partial \,^\mu 
\sigma +\partial _\mu \pi
^a\partial \,^\mu \pi ^a \right)\right. \nonumber \\ [3mm]
&&\left. -\frac 12\mu ^2\left(\left(\sigma+v \right)^2+
\pi ^a\pi ^a \right)
-\frac 1{4}\,\lambda \,
\left(\left(\sigma+v \right)^2+\pi ^a\pi ^a \right)^2 
\right. \nonumber \\ [3mm] 
&&\left. -g\,\bar{\psi}
\left(\left(\sigma+v \right)
+\,i\,\gamma ^5\tau ^a\pi ^a \right)\psi+c\,\sigma\right\}\,\,,  
\label {acao}
\end{eqnarray}}}
with $\gamma^\mu=e_m^\mu\gamma^m$, where $\gamma^m$ are the
Dirac matrices in the tangent space. 
g and $\lambda$ are the
pion-fermion and pion-pion coupling constants, respectively.
The covariant derivative is defined by 
\begin{equation}
{\cal D}_\mu \psi (x)= \left( \partial_\mu+\frac 12\omega 
_\mu ^{mn}(x)\Omega _{mn}\right) \psi (x)\,\,,
\end{equation}
with $\Omega^{[mn]}$ acting on $\psi$ as an infinitesimal 
Lorentz matrix in
the appropriate representation.

The masses arising from the action (\ref{acao}) for $\psi$, 
$\pi^a$ and $\sigma$ are given by:
\begin{eqnarray}
&&m_\psi =gv\,, \nonumber \\ [3mm]
&&m_\pi^2=\mu^2+\lambda v^2\,, \label {var} \\[3mm]
&&m_\sigma^2=\mu^2+3\lambda v^2\,\,. \nonumber
\end{eqnarray}

Moreover, from the Born approximation to the vaccum expectation
value~\cite{zuber}, we obtain
\begin{equation}
c=v(\mu^2+\lambda v^2)=v m_\pi^2\,\,. \label {var1}
\end{equation}

The isospin, chiral, diffeomorphisms and local Lorentz 
infinitesimal transformations
read:

{\it i}) Isospin
\begin{eqnarray}
\delta_{\rm iso} \,\sigma = 0\,\,\,,\,\,\,\delta_{\rm iso}
\,\psi =-\,i\,\frac{\alpha ^a\tau^a}
2\psi  \nonumber \\ [3mm]
\,\delta_{\rm iso} \,\pi ^a =\epsilon ^{abc}\alpha ^b\pi ^c\,\,
\,,\,\,\,\delta_{\rm iso} 
\,\bar{\psi}=i\,\bar{\psi}\,\frac{\alpha ^a\tau ^a}2\,\,.  
\label{iso1}
\end{eqnarray}

{\it ii}) Chiral
\begin{eqnarray}
\delta_{\rm chiral} \,\sigma =-\,\alpha ^a\pi ^a\,\,\,,\,\,\,
\delta_{\rm chiral} \,\psi =-\,i\,
\frac{\alpha ^a\tau ^a}2\gamma ^5\psi \,\,\,  \nonumber \\ [3mm]
\delta_{\rm chiral} \,\pi ^a =\alpha ^a(\sigma +v)\,\,\,,\,\,\,
\delta_{\rm chiral} \,\bar{\psi}=-i\,\bar{\psi}\gamma ^5\,
\frac{\alpha ^a\tau ^a}2\,\,,  \label{chiral1}
\end{eqnarray}

{\it iii}) Diffeomorphisms
\begin{eqnarray}
&&\delta _{{\rm diff}}^{(\varepsilon)}e_\mu^m ={\cal L}_\varepsilon e_\mu^m
=\varepsilon ^\lambda \partial _\lambda e_\mu^m +\left( \partial _\mu
\varepsilon ^\lambda \right) e_\lambda^m \,, \nonumber \\ [3mm]
&&\delta _{{\rm diff}}^{(\varepsilon )}\Phi ={\cal L}_\varepsilon \Phi
=\varepsilon ^\lambda \partial _\lambda \Phi \,,\quad \Phi =\sigma,
\,\pi^a,\,\psi\,\,, \label{diffeo0}
\end{eqnarray}
where ${\cal L}_\varepsilon $ is the Lie derivative along the vector field 
$\varepsilon ^{\mu}(x)$ -- the infinitesimal parameter of the transformation,

{\it iv}) Local Lorentz transformations
\begin{equation}
\delta_{{\rm Lorentz}}^{(\lambda)} \Phi = \frac12 \lambda_{mn}\Omega^{mn}
\Phi\,,\quad \Phi = \mbox{any field}\,\,,  \label{lor-fields}
\end{equation}
with infinitesimal parameters $\lambda_{[mn]}$.

The Ward identities corresponding to the isospin, chiral,
diffeomorphisms and local Lorentz symmetries for
the action (\ref{acao}) can be expressed introducing the functional
differential Ward operators, as given below:
\begin{eqnarray}
W_{\rm iso}^a\Sigma&=&\int d^4x\left(\epsilon ^{abc}\pi ^b\frac 
\delta {\delta
\pi ^c}+i\,\bar{\psi}\frac{\tau ^a}2\frac \delta 
{\delta \bar{\psi}} \right. \nonumber\\ [3mm]
&&\left. -\,i\frac{
\stackrel{\leftarrow }{\delta }}{\delta \psi }
\frac{\tau ^a}2\psi \right)
\Sigma =0\,,  \label{wiso}
\end{eqnarray}
\begin{eqnarray}
W_{\rm chiral}^a\Sigma&=&\int d^4x\left( (\sigma +v)\frac 
\delta {\delta \pi
^a}-\pi ^a\frac \delta {\delta \sigma}\right. \nonumber\\ [3mm]
&&\left. -\,i\,\bar{\psi}\gamma ^5\frac{\tau ^a}
2\frac \delta {\delta \bar{\psi}}
-i\frac{\stackrel{\leftarrow }{\delta }}{
\delta \psi }\frac{\tau ^a}2\gamma ^5\psi 
\right) \Sigma =\Delta
_{\rm class}^a\,, \nonumber \\ [3mm]
\label{wchiral}
\end{eqnarray}
where
\begin{equation}
\Delta _{\rm class}^a=-\int d^4x\,e\,c\,\pi ^a\,,  \label{quebra}
\end{equation}
is the breaking term which, being linear in the quantum field $\pi ^a$, will
not be renormalized, i.e., it will remain a classical breaking~\cite{pigsor},
\begin{equation}
{ W}_{{\rm diff}}\Sigma = \displaystyle{\int} d^3x\,\displaystyle{
\sum_{\Phi}^{}} \delta_{{\rm diff}}^{(\varepsilon)}\Phi {\displaystyle{\frac{
\delta\Sigma }{\delta\Phi }}} = 0 \,,  \label{diffeo}
\end{equation}
and 
\begin{equation}
{W}_{{\rm Lorentz}}\Sigma = \displaystyle{\int} d^3x\,\displaystyle{
\sum_{\Phi}^{}} \delta_{{\rm Lorentz}}^{(\lambda)}\Phi {\displaystyle{\frac{
\delta\Sigma }{\delta\Phi }}} = 0 \,,  \label{localoren}
\end{equation}
where the summations run over all quantum and external fields.

Finally, the classical action (\ref{acao}) is constrained, besides the Ward
identities (\ref{wiso}), (\ref{wchiral}), (\ref{diffeo}) and 
(\ref{localoren}) by a set of discrete symmetries,
i.e., parity $P$ and charge conjugation $C$, whose action on the
fields is given
as below:

{\it i}) Parity $P$: 
\begin{eqnarray}
&&x\stackrel{P}{\longrightarrow }\left( x^0,\,-\vec{x}\right) \,\,, 
\nonumber \\
&&\psi \stackrel{P}{\longrightarrow }\gamma ^0\psi \,\,,  \nonumber \\
&&\bar{\psi}\stackrel{P}{\longrightarrow }\bar{\psi}\gamma ^0\,\,,
\label{parity} \\
&&\pi ^a\stackrel{P}{\longrightarrow }\,-\,\pi ^a\,\,,  \nonumber \\
&&\sigma \stackrel{P}{\longrightarrow }\,\sigma\,\,.  \nonumber
\end{eqnarray}

{\it ii}) Charge conjugation $C$: 
\begin{eqnarray}
&&\psi \stackrel{C}{\longrightarrow }\psi ^c=C\,{\bar{\psi}}^t\,\,\,, 
\nonumber \\
&&\bar{\psi}\stackrel{C}{\longrightarrow }\bar{\psi}^c=
-\psi ^tC^{-1}\,\,, \nonumber \\
&&\pi ^{1,3}\stackrel{C}{\longrightarrow }\pi ^{1,3}\,\,,  \label{conj} \\
&&\pi ^2\stackrel{C}{\longrightarrow }-\pi ^2\,\,,  \nonumber \\
&&\sigma \stackrel{C}{\longrightarrow }\sigma \,\,,  \nonumber
\end{eqnarray}
where $C=i\,\gamma ^0\,\gamma ^2$ is the charge conjugation matrix.

Ultraviolet dimensions of all fields are collected in 
Table I~\footnote{The ultraviolet dimensions determine the 
ultraviolet power-counting.
If there are massless fields in the theory, one should take special
care of the infrared convergence~\cite{low}.}.


\section{Trace identity for the classical theory}

From now on we change from unphysical parametrization
$(\mu^2,\,\lambda,\,g,\,c)$ to physical one
$(m_\pi^2,\,m_\sigma^2,\,m_\psi,\,v)$, with the change of variables
given by (\ref{var}) and (\ref{var1}). In this way, the classical
action (\ref{acao}) takes the form:
\begin{eqnarray}
\Sigma &=&\int d^4x\,e\,\left\{ \bar{\psi}\,i\gamma ^\mu 
{\cal D}_\mu \psi
+\frac 12 \left(\partial _\mu \sigma \partial
\,^\mu \sigma +\partial _\mu \pi
^a\partial \,^\mu \pi ^a \right)\right. \nonumber\\ [3mm]
&&\left. -\,\frac {(3m_\pi^2-m_\sigma^2)}4 
\left(\left(\sigma+v \right)^2+\pi^a\pi^a \right) 
\right. \nonumber\\ [3mm]
&&\left. -\,\frac {(m_\sigma^2-m_\pi^2)}{8v^2}\,
\left(\left(\sigma+v \right)^2+\pi^a\pi^a \right)^2 
\right. \nonumber\\ [3mm]
&&\left. -\,\frac {m_\psi} {v}\,\bar{\psi}
\left(\left(\sigma+v \right) +\,i\gamma ^5\tau ^a\pi ^a \right)
\psi+m_\pi^2v
\,\sigma\right\}\,.  
\label {acao1}
\end{eqnarray}

For a given field theory the energy-momentum tensor is defined as the
functional derivative:
\begin{equation}
{\Theta }_\lambda ^{~\mu }= e^{-1} e_\lambda ^{~m}~\frac {\delta
\Sigma}{\delta e_\mu ^{~m}}\,\,.  \label{theta}
\end{equation}

The conservation of the energy-momentum tensor $\Theta _\lambda ^{~\mu }$ is
a consequence of the diffeomorphism Ward Identity (\ref{diffeo}) and of the
definition (\ref{theta}), yielding the following equation 
\begin{equation}
\int d^4x\,\varepsilon(x)\left(
e\,\nabla _\mu \Theta _\lambda ^{~\mu }\left( x\right)-w_\lambda \left(
x\right) \Sigma \right)=0 \,\,,
\label{a1}
\end{equation}
where $\nabla _\mu $ is the covariant derivative with respect to the
diffeomorphisms, with the differential operator $w_\lambda 
\left( x\right)$ acting on $\Sigma$ representing contact
terms: 
\begin{equation}
w_\lambda \left( x\right) =\sum\limits_
{\sigma,\,\pi,\,\psi}\left( \nabla
_\lambda \Phi \right) \frac \delta {\delta \Phi }\,\,,  \label{a9}
\end{equation}
(becoming the translation Ward operator in the limit of flat space).

The integral of the trace of the tensor $\Theta _\lambda ^{~\mu }$, 
\begin{equation}
\displaystyle{\int }d^4x\,e\,\Theta _\mu^{~\mu }=\displaystyle{\int}
d^4x\,e_\mu ^{~m}\frac{\delta \Sigma }{\delta e_\mu ^{~m}}
\equiv {\cal N}_e\Sigma \,\,,
\label{int-theta}
\end{equation}
turns out to be an equation of motion, up to soft breakings
 -- which means that $\Theta _\lambda
^{~\mu }$ is the improved energy-momentum tensor. This follows from the
identity, which is easily checked by inspection of the classical action 
\begin{eqnarray}
{\cal N}_e\Sigma&=&\left(\frac 32{\cal N}_\psi +{\cal N}_\Phi^{\rm hom}+
m_\sigma\,\frac \partial {\partial
m_\sigma}+m_\pi\,\frac \partial {\partial
m_\pi}\right. \nonumber\\ [3mm]
&&\left. +\,m_\psi\,\frac \partial {\partial
m_\psi}+v\,\frac \partial {\partial
v}\right)
\Sigma \,,  \label{int-theta1}
\end{eqnarray}
where ${\cal N}_e$ is the counting operator of the vierbein $e_\mu^m$.
${\cal N}_\psi$ and ${\cal N}_\Phi$ are counting operators defined by:
\begin{eqnarray}
{\cal N}_\psi &=&{\int}d^4x\,\,\left(\bar{\psi}\,\frac
\delta {\delta \bar{\psi}}+\frac{\stackrel{\leftarrow }{\delta }}
{\delta \psi }\psi \right)\,\,, \label {cop}
\\ [3mm]
{\cal N}_\Phi^{\rm hom} &=&{\int}d^4x\,\,\left(
\sigma\,\frac \delta {\delta \sigma
}+ \pi^a\,\frac \delta {\delta \pi^a}\right)\,\,. 
\label {cop1}
\end{eqnarray}
${\cal N}_\Phi^{\rm hom}$ is the unshifted counting operator
for scalar fields.

It is interesting to note that (\ref{int-theta1}) is nothing but the Ward
identity for rigid Weyl symmetry~\cite{iorio} -- broken by the mass terms and
dimensionful couplings.

Our classical ``trace identity'' is defined by:
\begin{equation}
W^{\rm trace}\Sigma=\Lambda\,\,, \label{tr}
\end{equation}
where
\begin{equation}
W^{\rm trace}={\cal N}_e-\frac 32{\cal N}_\psi-
{\cal N}_\Phi^{\rm hom}\,\,, \label{tr1}
\end{equation}
and
\begin{eqnarray}
\Lambda &=&\int d^4x\,e\,\left(-m_\sigma^2\sigma^2-m_\pi^2\pi^a\pi^a
-m_\psi\bar{\psi}\psi \right. \nonumber\\ [3mm]
&&\left. -\,\frac {(m_\sigma^2-m_\pi^2)}{2v}\,\sigma
\left(\sigma^2+\pi^a\pi^a\right)\right)\,.  
\label {bre}
\end{eqnarray}

The latter is the effect of the breaking of scale invariance due to the
dimensionful parameters. The dimension of $\Lambda$ -- the dimensions
of $m_\pi^2,\,m_\sigma^2,\,m_\psi$ and $v$ not being taken into account
-- is lower than four: it is a soft breaking.

In order to make the connection between the trace identity and the
dilatational Ward identity -- which is just the scaling
Ward identity --
let us consider a while the limit of flat space, where rigid
dilatation symmetry makes sense. In this limit (\ref{a1}) holds
with $e=1$ and $\nabla_\mu=\partial_\mu$. The classical dilatation current
${\cal S}^\mu$ can now be defined as
\begin{equation}
{\cal S}^\mu \left( x\right)=x^\lambda\, \Theta _\lambda ^{~\mu }\left(
x\right)\,\,.  \label{dilata}
\end{equation}

It obeys the conservation identity
\begin{eqnarray}
\partial _\mu {\cal S}^\mu&=&\Theta_\mu^{~\mu}+
x^\lambda\, \partial_\mu \Theta_\lambda^{~\mu} \nonumber \\ [3mm]
&=&\left(\frac 32 +x\cdot\partial\right)n_\psi \Sigma+
\left(1+x\cdot\partial\right)n_\Phi^{\rm hom} \Sigma \nonumber\\ [3mm]
&&+\,\Lambda(x)\,\,,
\label{dilata1}
\end{eqnarray}
where $\Lambda(x)$ is the integrand of (\ref{bre}).
$n_\psi$ and
$n_\Phi^{\rm hom}$ are local operators given by
\begin{eqnarray}
n_\psi\left(x\right) &=&\left(\bar{\psi}\,\frac
\delta {\delta \bar{\psi}}+\frac{\stackrel{\leftarrow }{\delta }}
{\delta \psi } \psi \right)\,\,, \nonumber \\ [3mm]
n_\Phi^{\rm hom}\left(x\right) &=&\left(
\sigma\,\frac \delta {\delta \sigma
}+ \pi^a\,\frac \delta {\delta \pi^a}\right)\,\,. 
\nonumber
\end{eqnarray}

Integrating (\ref{dilata1}) we get the broken dilatation Ward identity
for the
classical theory: 
\begin{equation}
W_{\rm D} \Sigma= \int d^4x \sum\limits_
{{\Phi=\sigma,\,\pi^a,\,\psi}}\left[ \left( d_\Phi
+x\cdot \partial \right) \Phi \right] \frac {\delta \Sigma} 
{\delta \Phi}
=\Lambda\,\,,
\label{a17'}
\end{equation}
where $d_\Phi$ is the
dimension of the field $\Phi$ (see Table 1).

The trace Ward operator defined in (\ref{tr}), together with the
operators 
(\ref{wiso}), (\ref{wchiral}),
(\ref{diffeo}) and (\ref{localoren}) fulfill the algebra
\begin{eqnarray}
\left[W^{\rm trace},\,W_{\rm chiral}^a\right]\,{\cal F}&=&
\int d^4x\,\,v\,\frac {\delta {\cal F}} {\delta \pi^a}\,\,,
\label {relcom} \\ [3mm]
\left[W^{\rm trace},\,W_{\rm X}\right]\,{\cal F}&=&0\,\,\,,
\,\,\,{\rm X}={\rm iso},\,{\rm Lorentz},\,{\rm diff}\,\,,
\label {relcom1}
\end{eqnarray}
where ${\cal F}$ is an arbitrary functional.

\section{Trace identity for the quantum theory}

We have now to extend the construction of the preceding Section
to all orders of perturbation theory. As a starting point, we must
be able to write the breaking of scale invariance (\ref{bre}) in form
of a differential operator. This will be possible with the help 
of additional
external fields $\eta$ and $\rho$, which transform invariantly
under isospin and chiral symmetries -- $P$ and $C$-even -- 
introducing the
new classical action
\begin{equation}
{\Sigma}^\natural=\Sigma+
\int d^4x\,e\,\left(a\,\rho{\cal Q}^{\rm inv}
+3\,m_\pi^2v\eta\sigma\right)\,\,, \label{newacao1}
\end{equation} 
with
\begin{eqnarray}
{\cal Q}^{\rm inv}&=&\left(\sigma+v\right)^2+\pi^a\pi^a\,\,,
\nonumber 
\end{eqnarray}
an invariant polynomial of dimension two. The ultraviolet 
dimension of $\eta$ and $\rho$ is 2.

Following along the lines of~\cite{kraus1}, note that
\begin{eqnarray}
&&\left[\left(\int d^4x\,v\,\frac{\delta} {\delta \sigma}+
\frac{\delta} {\delta \eta}+a\,\frac{\delta} {\delta \rho}
\right),\,W_{\rm chiral}^a\right]\,{\cal F}= \nonumber\\ [3mm]
&&\int d^4x\,\,v\,\frac {\delta {\cal F}} {\delta \pi^a}\,\,.
\label {relcom12}
\end{eqnarray}

Taking into account (\ref{relcom}) and (\ref{relcom12}),
for ${\cal F}=\Sigma^\natural$ (setting at the end $\eta=\rho=0$),
one finds:
\begin{eqnarray}
&&W_{\rm chiral}^aW^{\rm trace}\left.
{\Sigma}^\natural\right|_{\eta=\rho=0}= \nonumber\\ [3mm]
&&W_{\rm chiral}^a\left(\int d^4x\,\left(v\,
\frac {\delta} {\delta \sigma}
+\frac \delta {\delta\eta}
+a \frac \delta {\delta \rho}\right)\left.
{\Sigma}^\natural\right|_{\eta=\rho=0}\right)= \nonumber\\ [3mm]
&&W_{\rm chiral}^a\Lambda\,\,.
\label {relcom41}
\end{eqnarray}

Therefore, in the tree approximation one gets the expression:
\begin{eqnarray}
\Lambda=
\int d^4x\,\left(v\,\frac {\delta} {\delta \sigma}
+\frac \delta {\delta\eta}
+a \frac \delta {\delta \rho}\right)\left.
{\Sigma}^\natural\right|_{\eta=\rho=0}\,\,,
\label {relcom4}
\end{eqnarray}
with
\begin{equation}
a= \frac {(m_\sigma^2-3m_\pi^2)} {2}\,\,,
\end{equation}
determined by the normalization conditions.

We now define the ``symmetrized'' form of the classical trace identity 
\begin{eqnarray}
&&\widehat{W}^{\rm trace}\left.
{\Sigma}^\natural\right|_{\eta=\rho=0}
= \nonumber\\ [3mm]
&&\left(W^{\rm trace}-
\int d^4x\,\left(v\,\frac {\delta} {\delta \sigma}
+ \frac \delta {\delta\eta}
+a \frac \delta {\delta \rho}\right)\right)
\left.
{\Sigma}^\natural\right|_{\eta=\rho=0}=0
\,\,. \nonumber\\
\label {newtrop}
\end{eqnarray}

With the new operator $\widehat{W}^{\rm trace}$, we have:
\begin{eqnarray}
\left[\widehat{W}^{\rm trace},\,{W}_{\rm chiral}^a\right]
\,{\cal F}&=&0\,\,,
\nonumber \\ [3mm]
\left[\widehat{W}^{\rm trace},\,W_{\rm X}\right]\,{\cal F}&=&0\,\,.
\label {relcom6}
\end{eqnarray}
where ${\rm X}={\rm iso},\,{\rm Lorentz},\,{\rm diff}$.

Let us note that
\begin{eqnarray}
W_{\rm chiral}^a{\Sigma}^\natural={\Delta}_{\rm class}^{\natural\,a}\,\,,
\end{eqnarray}
where
\begin{equation}
{\Delta}_{\rm class}^{\natural\,a}=-\int d^4x\,e\,\left(
1+3\,\eta\right)m_\pi^2v\pi^a\,\,,  \label{quebra505}
\end{equation}
is a breaking term which stays linear in the quantum field 
$\pi^a$, and
will remain a classical breaking~\cite{pigsor}.
Thus, the proof of the renormalizability for
(\ref{newacao1}) remains the same as the one sketched in appendix A.

The corresponding quantum theory is described introducing the 
new vertex functional
\begin{equation}
{\Gamma}^\natural={\Sigma}^\natural+O(\hbar)\,\,. \label{newvertfu}
\end{equation}
In this way, the trace identity (\ref{newtrop}) takes the form
\begin{eqnarray}
\widehat{W}^{\rm trace}\left.
{\Gamma}^\natural\right|_{\eta=\rho=0}
=\Delta\cdot
\left.
{\Gamma}^\natural\right|_{\eta=\rho=0}=
\Delta+O(\hbar)
\,\,.
\label {newtrop12}
\end{eqnarray}
The insertion in the right-hand side represents the breaking
due to the effect of
the radiative corrections we want to study, and $\Delta$ is their
lowest order contribution. From Quantum Action Principle (QAP)~\cite{qap},
$\Delta$ is an integrated field polynomial compatible with ultraviolet
dimension 4 and even under the parity $P$ and charge conjugation $C$.

According also to the QAP, applying
the algebraic structure (\ref{relcom6}) to the vertex functional,
one gets:
\begin{eqnarray}
{W}_{\rm chiral}^a\widehat{W}^{\rm trace}
\,\left.\Gamma^\natural \right|_{\eta=\rho=0}&=&
{W}_{\rm chiral}^a \Delta=0\,\,,
\nonumber \\ [3mm]
W_{\rm X}\widehat{W}^{\rm trace}
\,\left.\Gamma^\natural \right|_{\eta=\rho=0}&=&
W_{\rm X} \Delta=0
\,\,.
\label {relcom61}
\end{eqnarray}
with ${\rm X}={\rm iso},\,{\rm Lorentz},\,{\rm diff}$.

For that reason, $\Delta$ is
an invariant
insertion which can be expanded in a suitable basis. It is
remarkable that, in perturbation theory, any such basis of
renormalized insertions is completely characterized by the
corresponding classical basis~\cite{pigsor}. Such a basis
is given in the classical approximation by (\ref{poly1})
-- see appendix A. An appropriate quantum extension of this
basis is obtained through the introduction
of a set of symmetric differential operators acting on
$\Gamma^\natural$ -- setting at the end $\eta=\rho=0$ --
and in one-to-one correspondence to the basis of integrated
polynomials in (\ref{poly1}). We define a symmetric operator as an operator
$\nabla$ which fulfills the condition
\begin{equation}
\left[\nabla,\,W_{\rm X}\right]=0\,\,\,,\,\,\,{\rm X}={\rm iso},\,
{\rm chiral},\,{\rm Lorentz},\,{\rm diff}\,\,. \label{sym}
\end{equation} 

The set
\begin{equation}
\left\{ \int d^4x\,a\,\frac \delta {\delta
\rho},
\,v \frac \partial {\partial v
},
\,m_\psi\frac \partial {\partial m_\psi},
\,{\cal N}_\Phi,\,
{\cal N}_\psi 
\right\}\,\,,  \label{counter55}
\end{equation}
with ${\cal N}_\psi$ given by (\ref{cop}) and
\begin{eqnarray}
{\cal N}_\Phi &=&{\int}d^4x\,\,\left((
\sigma+v)\,\frac \delta {\delta \sigma
}+ \pi^a\,\frac \delta {\delta \pi^a}\right)
\label{cop2}\,\,, 
\end{eqnarray}
forms a basis for the symmetric operators of the model, taking into
account the physical parametrization.

Thus, the expansion of $\Delta$
in the basis (\ref{counter55}), we have just constructed, yields
the quantum trace identity in the curved space-time:
\begin{eqnarray}
&&\int d^4x\,e\,\Theta_\mu^{~\mu} \cdot
\left.\Gamma^\natural \right|_{\eta=\rho=0}=
\left(\beta_{m_\psi}\,m_\psi\frac \partial {\partial m_\psi}
+\beta_v\,v \frac \partial {\partial v
} \right. \nonumber\\ [3mm]
&&\left.+\left(\frac 32-\gamma_\psi\right){\cal N}_\psi
+\left(1-\gamma_\Phi\right){\cal N}_\Phi^{\rm hom}
\right. \nonumber \\
[3mm]
&&\left.+\left(1-\beta_v-\gamma_\Phi\right)\int d^4x\,v\, \frac \delta
{\delta \sigma}+\left(1+\delta\right)\int d^4x\,a\,\frac \delta
{\delta \rho} \right. \nonumber \\ [3mm]
&&\left.+\left(1+\beta_v\right) \int d^4x\, \frac
\delta {\delta \eta} \right) \left.\Gamma^\natural
\right|_{\eta=\rho=0}
\nonumber \\ [3mm]
&&+\,{\rm terms\,\,vanishing\,\,in\,\,the\,\,flat\,\,limit}\,\,,
\label{qbre} 
\end{eqnarray}
where ${\cal N}_\Phi^{\rm hom}$ is an unshifted counting operator.  
 
In the flat space, (\ref{qbre}) is equivalent to the Callan-Symanzik
equation, which is the Ward identity for anomalous dilatation
invariance -- see appendix B. It is worthwhile to note that this
result allows us
to conclude which the $\beta$-functions and anomalous dimensions
in curved space are the same as in flat space. The presence of
the $\beta_{m_\psi}$-function corresponds
to renormalization of the physical mass of fermionic fields,
with the consequence that the hard breaking of dilatations depends 
on the normalization point also in the asymptotic region. 

\section{Summary}

\setcounter{equation}{0} 

In this paper, we show -- by using the techniques developed
in~\cite{odhp} -- as to identify directly the breaking of dilatations
with the trace of the energy-momentum tensor in a model with
explicitly broken symmetry. This is not a trivial task due the
shifts by
constant amounts in certain fields: the dilatational operator
does not commute with the Ward operator for broken symmetry,
but has a certain covariance under the symmetry transformations
already at the classical level. Most remarkable is the
presence of the $\beta$-function associated with the physical
mass of fermions. According Becchi~\cite{becchi}, a ``true''
CS equation
does not exist in such situation. By ``true'' it shoud be understood
an equation which does not contain $\beta$-functions belonging
to the physical mass differential operators. In any case this requires
an analysis. This effort is essential if one aims
at having contact with phenomenology. In particular, this is
the case for a realistic
supersymmetric gauge field theory \cite{mpw}.
The reader may convince himself that our algorithm also
works for the case of spontaneous symmetry breaking.
In this case, due to the
eventual appearance of Goldstone modes, infrared anomalies may
be picked up, and in higher order have to be proven to be absent
\cite{kraus1}.
As a by-product, the approach has allowed us to conclude 
that the
$\beta$-functions and the anomalous dimensions in curved space are
the same as in flat space -- evidently this is valid for a class
of curved manifolds with topology remains that of flat ${\cal R}^4$
and with asymptotically vanishing curvature. It is only in this
case we can use the general results of renormalization theory,
established in flat space.

{\bf Acknowledgements:}
I am deeply indebted
to Prof. O. Piguet for helpful comments on a preliminary draft of
this paper.
I wish to thank Prof. L. Bonora for his kind
invitations at the Scuola Internazionale Superiore di Studi Avanzati
(SISSA) in Trieste, Italy. Thanks are also due to J.A. Helay\"el-Neto,
E.A. Pereira and O.M. Del Cima for encouragement.



\appendix 
\renewcommand{\theequation}{\Alph{section}.\arabic{equation}} 



\section{Algebraic proof of renormalizability}


In this appendix, we sketch a proof of renormalizability of the 
chiral fermion
meson model on the light of the regularization-independent algebraic
method\footnote{ In fact, this has already 
been considered in~\cite{becchi} via BPHZ renormalization scheme and
recently in
\cite{dhac} via ``algebraic'' renormalization for the theory in flat 
space only. The generalization to curved space is
straightforward.}.

In the first step, we
study the stability of the classical action. For the quantum theory the
stability corresponds to the fact that the radiative corrections can be
reabsorbed by a redefinition of the initial parameters of the theory. Next,
one computes the possible anomalies through an analysis of the Wess-Zumino
condition, then one checks if the possible breakings induced by
radiative corrections can be fine-tuned by a suitable choice of
non-invariant local counterterms.

\subsection{Stability}

In order to study the stability of the model under radiative corrections, we
introduce an infinitesimal perturbation in the classical action $\Sigma $ by
means of a integrated local functional $\tilde{\Sigma}$ that satisfies the
constraint of a quantum correction
\begin{equation}
\Sigma \rightarrow \Sigma +\epsilon \,\tilde{\Sigma},  \label{stabe}
\end{equation}
where $\epsilon $ is an infinitesimal parameter.

The perturbed action must satisfy, to the order $\epsilon $, the same
equations as $\Sigma$, i.e.:
{\small{
\begin{eqnarray}
W_{\rm X}\left( \Sigma +\epsilon \,\tilde{\Sigma}\right)
&=&W_{\rm X}\left( \Sigma \right) +\epsilon \,
W_{\rm X}\tilde{\Sigma}
+O\left( \epsilon ^2\right) =0\,\,,\nonumber \\ [3mm]
W_{\rm iso}^a\left( \Sigma +\epsilon \,\tilde{\Sigma}\right)
&=&W_{\rm iso}^a\left( \Sigma \right) +\epsilon \,
W_{\rm iso}^a\tilde{\Sigma}
+O\left( \epsilon ^2\right) =0\,\,, \nonumber\\ [3mm]
W_{\rm chiral}^a\left( \Sigma +\epsilon \,\tilde{\Sigma}\right)
&=&W_{\rm chiral}^a\left( \Sigma \right) +\epsilon \,
W_{\rm chiral}^a\tilde{\Sigma}
+O\left( \epsilon ^2\right) \nonumber\\ [3mm]
&=&\Delta_{\rm class}^a\,\,.
\label{acaope}
\end{eqnarray}}}
with ${\rm X}={\rm Lorentz},\,{\rm diff}$.

To first order in $\epsilon$ , one obtains:
\begin{equation}
W_{\rm X}\tilde{\Sigma}=0\,\,\,,\,\,\,
W_{\rm iso}^a\tilde{\Sigma}=0\,\,\,,
\,\,\,W_{\rm chiral}^a\tilde{
\Sigma}=0\,\,\,,  \label{acaope1}
\end{equation}
consequently all counterterms required by renormalization have to be
symmetric.

Let us look for the most general invariant counterterm $\tilde{\Sigma}$ ,
i.e., the most general field polynomial of UV dimension $\leq $ 4,
respecting parity and charge conjugation symmetries and the conditions
(\ref{acaope1}). An explicit computation, shows that $\tilde{\Sigma}$
can be written in the following way:
\begin{eqnarray}
\tilde{\Sigma}=\int d^4\,e\,\sum\limits_{i=1}^5a_i{\cal P}_i(x)
\,\,, \label{poly1}
\end{eqnarray}
where 
\begin{eqnarray}
&&{\cal P}_1=\bar{\psi}\,i\gamma ^\mu {\cal D}_\mu \psi\,,\,\,
{\cal P}_2=\left( \partial _\mu \sigma \partial \,^\mu \sigma +\partial_\mu 
\pi^a\partial \,^\mu \pi ^a\right)\,, \nonumber \\ [3mm]
&&{\cal P}_3=\left(
(\sigma +v)^2+\pi
^a\pi ^a\right)\,,\,\,
{\cal P}_4=\left((\sigma
+v)^2+\pi ^a\pi ^a\right)^2\,, \nonumber \\ [3mm]
&&{\cal P}_5=\bar{\psi}\left((\sigma
+v)+\,i\,\gamma ^5\tau ^a\pi^a\right) \psi \,\,,  \nonumber
\end{eqnarray}
with $a_1,...,a_5$ arbitrary coefficients. We have negleted terms
such as $\int d^4\,e\,R\,(\sigma^2+\pi^a\pi^a)$, which do not
contribute in the limit of flat space. 

The arbitrary coefficients are fixed in such a way
that they hold order by order in perturbation theory by normalization 
conditions. Considering the physical parametrization, adopted in the
main text, and since we will have a particle interpretation only if the
vaccum expectation value of the fields vanish, we impose the following
non-singular system of normalization conditions:
\begin{eqnarray}
&&\Gamma _{\sigma}=0\,\,\,, 
\,\,\, \left. \frac \partial {\partial\,p^2}
\Gamma_{\sigma \sigma}\right| _{p^2=\kappa}=1
\,\,\,,\,\,\,
\left. \frac \partial {\partial \,\rlap
{\hbox{$\mskip 1 mu /$}}p}
\Gamma_{
\bar{\psi}\psi }\right| _{\not{p}=\kappa}=1\,\,,
\nonumber \\ [3mm]
&&\left. \Gamma_{\sigma \sigma}
\right|_{p^2=m_\sigma^2}
=0\,\,\,,\,\,\,
\left. \Gamma _{\bar{\psi}\psi}\right|_{\not{p}=m_\psi}
=0\,\,.  \label{recond20}
\end{eqnarray}
where $\kappa $ is a energy scale and $p\left( \kappa \right)$ some
reference set of $4$-momenta at this scale. $m_\pi^2$ is determined
through the chiral Ward identity (\ref{wchiral}). With the normalization
conditions (\ref{recond20}), the most general action becomes 
identical to the action
(\ref{acao1}).

\subsection{Anomalies}

Because the classical stability does not imply in general the possibility
of extending the theory to the quantum level, our second task is to
infer for possible anomalies. Then, a generating functional for vertex
functions, $\Gamma$, is constructed
\begin{equation}
\Gamma =\Sigma +O\left( \hbar \right) ,  \label{funvert}
\end{equation}
such that
\begin{equation}
W_{\rm X}\Gamma=0\,\,,\,\,\,W_{\rm iso}^a\Gamma =0\,\,,\,\,\,
W_{\rm chiral}^a\Gamma
=\Delta _{\rm class}^a\,\,.  \label{acaope2}
\end{equation}

The validity to all orders of the Ward identities
of diffeomorphisms and local Lorentz will be assumed in the following.
In fact, in the absence of gauge fields, these anomalies can exist 
only in $D=4k+2$ dimensions, with
$(k=0,1,2,...)$, represented
by a local polynomial in the curvature only
(see~\cite{bertlmann} and references cited therein)\footnote{See 
{\it e.g.}~\cite{bonora}
where the authors study the cohomology problem of the overall local
symmetry group of theories with external gravity, including
diffeomorphisms, local Lorentz and gauge transformations, in order
to determine all possible anomalies.}.

It remains now to show the possibility of implementing the isospin and
chiral Ward identities for the vertex functional $\Gamma$.
The proof is recursive. We shall admit the assumption that there exists a
vertex functional $\Gamma ^{\left( n-1\right) }$ obeying the 
Ward identities
(\ref{acaope2}) till the order $n-1$ in $\hbar$,
\begin{eqnarray}
W_{\rm iso}^a\Gamma^{\left( n-1\right) }&=&O\left( \hbar ^n\right) \,\,,
\label{form1} \\ [3mm]
\,W_{\rm chiral}^a\Gamma ^{\left( n-1\right) }&=&\Delta _{\rm class}
^a+O\left( \hbar
^n\right) \,\,\,.  \label{form2}
\end{eqnarray}

As a result of QAP~{\cite{qap}}, the forms (\ref
{form1}) e (\ref{form2}) will be broken at the $n$-order as follows
\begin{equation}
W_{\rm iso}^a\Gamma ^{\left( n-1\right) }=\hbar ^n\Delta \cdot \Gamma =\hbar
^n\Delta _{\rm iso}^a+O\left( \hbar ^{n+1}\right) \,\,,  \label{anom}
\end{equation}
\begin{eqnarray}
W_{\rm chiral}^a\Gamma ^{\left( n-1\right) }&=&\Delta _{\rm class}
^a+\hbar ^n\Delta
\cdot \Gamma \nonumber\\ [3mm] 
&=&\Delta _{\rm class}^a+\hbar^n\Delta _{\rm chiral}
^a+O\left( \hbar
^{n+1}\right) \,\,\,,  \label{wquiral}
\end{eqnarray}
where $\Delta _{\rm iso}^a$ and $\,\Delta _{\rm chiral}^a$,
are integrated local
functionals with UV\thinspace dimension $\leq$ 4.

Due to the invariance under parity $P$ and charge conjugation $C$, the Ward
operators and the quantum breakings satisfy the properties

{\it i}) Parity $P$
\begin{eqnarray}
&&W_{\rm iso}^a\stackrel{P}{\longrightarrow }W_{\rm iso}^a\,\,,\,\,\,\,\,
\,\,\,\,
\,\,\,\,\,\,\,\,\,\,\,\Delta _{\rm iso}^a\stackrel{P}
{\longrightarrow }\Delta
_{\rm iso}^a\,\,,  \nonumber \\ [3mm]
&&W_{\rm chiral}^a\stackrel{P}{\longrightarrow }-W_{\rm chiral}^a
\,\,,\,\,\,\,\,\,\,\,
\,\,\,\Delta _{\rm chiral}^a\stackrel{P}{\longrightarrow}
-\Delta _{\rm chiral}^a\,\,,
\label{parity1}
\end{eqnarray}

{\it ii}) Charge conjugation $C$
\begin{eqnarray}
&&W_{\rm iso}^{1,3}\stackrel{C}{\longrightarrow }-
W_{\rm iso.}^{1,3}\,\,,\,\,\,\,\,\,\,
\Delta _{\rm iso}^{1,3}\stackrel{C}{\longrightarrow }-\Delta
_{\rm iso}^{1,3}\,\,,  \nonumber \\ [3mm]
&&W_{\rm iso}^2\stackrel{C}{\longrightarrow }
W_{\rm iso}^2\,\,,\,\,\,\,\,\,\,\,\,
\,\,\,\,\Delta _{\rm iso}^2\stackrel{C}
{\longrightarrow }\Delta _{\rm iso}^2\,\,, 
\nonumber \\ [3mm]
&&W_{\rm chiral}^{1,3}\stackrel{C}{\longrightarrow}W
_{\rm chiral}^{1,3}\,\,,\,\,\,\,
\,\,\,\,\,\Delta _{\rm chiral}^{1,3}\stackrel{C}{\longrightarrow }\Delta
_{\rm chiral}^{1,3}\,\,,  \nonumber \\ [3mm]
&&W_{\rm chiral}^2\stackrel{C}{\longrightarrow }-W_{\rm chiral}^2
\,\,,\,\,\,\,\,\Delta
_{\rm chiral}^2\stackrel{C}{\longrightarrow}-\Delta _{\rm chiral}^2\,\,,
\label{conj1}
\end{eqnarray}

Using the commutation relation $\left[ \tau ^{a},\,\tau ^{b} \right]
=2\,i\,\epsilon ^{abc} \tau ^{c}$, it is easy to check that the Ward
operators obey the following commutation rules of the Lie algebra:
\begin{eqnarray}
\left[ W_{\rm iso}^a,W_{\rm iso}^b\right]{\cal F} &=&-\,\epsilon^
{abc}W_{\rm iso}^c {\cal F}\,\,, 
\nonumber \\ [3mm]
\left[ W_{\rm iso}^a,W_{\rm chiral}^b\right]{\cal F} &=&-\,\epsilon^{abc}
W_{\rm chiral}^c {\cal F}\,\,,
\label{algebra} \\ [3mm]
\left[ W_{\rm chiral}^a,W_{\rm chiral}^b\right]{\cal F} &=&-\,\epsilon^
{abc}W_{\rm iso}^c {\cal F}\,\,, 
\nonumber
\end{eqnarray}
with $\cal F$ an arbitrary functional.

Applying the algebraic structure above displayed to the vertex functional,
we obtain the Wess-Zumino consistency conditions~{\cite{wezu}}\thinspace
\thinspace \thinspace \thinspace \thinspace \thinspace \thinspace 
\begin{eqnarray}
W_{\rm iso}^a\Delta _{\rm iso}^b-W_{\rm iso}^b\Delta _{\rm iso}^a 
&=&-\epsilon
^{abc}\Delta _{\rm iso}^c\,\,,  \nonumber \\ [3mm]
W_{\rm iso}^a\Delta _{\rm chiral}^b-W_{\rm chiral}^b\Delta _{\rm iso}^a 
&=&-\epsilon
^{abc}\Delta _{\rm chiral}^c\,\,,  \label{wess} \\ [3mm]
W_{\rm chiral}^a\Delta _{\rm chiral}^b-W_{\rm chiral}^b\Delta
_{\rm chiral}^a &=&-\epsilon
^{abc}\Delta _{\rm iso}^c\,\,.  \nonumber
\end{eqnarray}

Solving constraints such as (\ref{wess}) is technically known as a problem
of Lie algebra cohomology. Its solution can always be written as a sum of a
trivial cocycle $W_{\rm iso(chiral)}^a\Delta $, and of nontrivial elements
belonging to the cohomology of $W_{\rm iso(chiral)}^a$:
\begin{equation}
\Delta _{\rm iso(chiral)}^a={\cal A}_{\rm iso(chiral)}^a+
W_{\rm iso(chiral)}^a\Delta
\,\,.  \label{anom1}
\end{equation}

As it is well know the theory will be anomaly free if 
conditions (\ref{wess}) admit only the trivial solution
\begin{equation}
\Delta _{\rm iso(chiral)}^a=W_{\rm iso(chiral)}^a\Delta \,\,\,,
\end{equation}
with $\Delta $ an integrated local functional even under parity and charge
conjugation. On the other hand, non-trivial cocycles, i.e,
\begin{equation}
{\cal A}_{\rm iso(chiral)}^a\neq W_{\rm iso(chiral)}^a\Delta \,\,,
\end{equation}
cannot be reabsorbed as local counterterms and represent an obstruction in
order to have an invariant quantum vertex functional.

A direct inspection shows that there is no such a functional satisfying
(\ref{parity1}) and (\ref{conj1}) for $\Delta _{\rm iso}^a$. Hence, its
cohomology is empty. On the other hand, the chiral
breaking, $\Delta _{\rm chiral}^a$, can be expanded in the basis: 
{\small{
\begin{eqnarray}
&&{\int}d^4x\,e\,\left(\pi^a\,,\,\pi^a\sigma ^2\,,\,\pi ^a\sigma ^3\,,\, 
\partial _\mu
\pi ^a\partial ^\mu \sigma\,,
\bar{\psi}\gamma ^5\tau ^a\psi\,,\,
\pi^a\bar{\psi}\psi \right)\,\,. \nonumber\\
\label{poly2}
\end{eqnarray}}}
parity and charge conjugation being taken into account.

The consistency conditions (\ref{wess}) reduce this basis to
\begin{equation}
{\int}d^4x\,e\,\left(\,\pi^a\,,\,\bar{\psi}\gamma^5\tau^a\psi
\,\right) \,\,.
\end{equation}

Such a basis can be obtained by applying $W_{\rm chiral}^a$ to
\begin{equation}
{\int}d^4x\,e\,\left(\,\sigma\,,\,{\bar{\psi}\psi}\,\right)\,\,,
\end{equation}
i.e., it can be reabsorbed as local counterterms.

Denoting the latter field monomials by $\Delta _i$, we can write (\ref
{wquiral}) as
\begin{eqnarray}
W_{\rm chiral}^a\Gamma ^{\left( n-1\right) }&=&\Delta _{\rm class}
^a+\hbar ^n\Delta
\cdot \Gamma \nonumber\\ [3mm] 
&=&\Delta _{\rm class}^a+\hbar ^nW_{\rm chiral}^a\Delta 
+O\left( \hbar
^{n+1}\right) \,\,, \nonumber\\ [3mm]
\end{eqnarray}
where $\Delta =\sum\limits_{i=1}^2r^i\Delta _i$.

Defining $\Sigma^{\left( n-1\right) }$ the action, with all its 
couterterms till
the order $n-1$, which leads to the functional $\Gamma^{\left( n-1\right)}$,
then replacing the action $\Sigma^{\left( n-1\right) }$ by the new action
\begin{equation}
\Sigma^{\left( n\right) }=\Sigma^{\left( n-1\right) }-\hbar ^n\Delta \,\,,
\end{equation}
lead to the new vertex functional
\begin{equation}
\Gamma ^{\left( n\right) }=\Gamma ^{\left( n-1\right) }-\hbar ^n\Delta
\,+O\left( \hbar ^{n+1}\right) \,\,\,.
\end{equation}

Thus, with the results obtained above, we get
\begin{equation}
W_{\rm iso}^a\Gamma ^{\left( n\right) }=O\left( \hbar ^{n+1}\right) \,\,,\,
\end{equation}
\begin{equation}
W_{\rm chiral}^a\Gamma ^{\left( n\right) }=\Delta _{\rm class}
^a+O\left( \hbar
^{n+1}\right) \,\,\,,
\end{equation}
which is the next order Ward identities we wanted to prove.


\section{Callan-Symanzik equation}


In this appendix we wish to derive the CS equation. This allows
us to identify the coefficients $\beta$ and $\gamma$ of the
expression (\ref{qbre}) with those of the CS equation, when
we take the limit of flat space-time. Our starting point is
eq.(\ref{a17'}), the broken dilatation Ward identity, with
$\Sigma$ replaced by
$\Sigma^\natural$
{\small{
\begin{eqnarray}
W_{\rm D} \left.\Sigma^\natural\right|_{\eta=\rho=0}&=& 
\int d^4x \sum\limits_
{{\Phi=\sigma,\,\pi^a,\,\psi}}\left[ \left( d_\Phi
+x\cdot \partial \right) \Phi \right] 
\left.\frac {\delta \Sigma^\natural} 
{\delta \Phi}\right|_{\eta=\rho=0}\nonumber \\ [3mm]
&=&\Lambda\,\,.
\label{a17b}
\end{eqnarray}
}}

With the help of eq.(\ref{relcom4}), we can define the ``symmetrized''
form of (\ref{a17b})
{\small{
\begin{eqnarray}
\left(W_{\rm D}-
\int d^4x\,\left(v\,\frac {\delta} {\delta \sigma}
+\frac \delta {\delta\eta}
+a \frac \delta {\delta \rho}\right)\right)\left.
{\Sigma}^\natural\right|_{\eta=\rho=0}=0\,\,.
\nonumber\\ \label {relcom4a}
\end{eqnarray}
}}

Applying QAP one derives
from (\ref{relcom4a}) that the dilatations in higher order,
are broken by
\begin{eqnarray}
&&\left(W_{\rm D}-
\int d^4x\,\left(v\,\frac {\delta} {\delta \sigma}
+\frac \delta {\delta\eta}
+a \frac \delta {\delta \rho}\right)\right)\left.
{\Gamma}^\natural\right|_{\eta=\rho=0}= \nonumber\\ [3mm]
&&\Delta \cdot \left.
{\Gamma}^\natural\right|_{\eta=\rho=0}=
\Delta + O(\hbar)\,\,,
\label {relcom4b}
\end{eqnarray}
where $\Gamma^\natural$ is the vertex functional defined in
(\ref{newvertfu}). $\Delta$ represents the breaking in the
lowest order.

According to the fact that the left-hand side of (\ref{relcom4b})
it is symmetric with respect to isospin and chiral symmetries,
one gets: 
\begin{eqnarray}
W_{\rm iso}^a\Delta=W_{\rm chiral}^a\Delta=0
\,\,.
\label {relcom4c}
\end{eqnarray}

The invariant insertion $\Delta$ can be expanded in a suitable
basis of symmetric operators of the theory -- parity and charge
conjugation being taken into account. Assuming the physical
parametrization, this basis is given by set of operators
(\ref{counter55}), yielding
\begin{eqnarray}
&&\left(W_{\rm D}-
\int d^4x\,\left(v\,\frac {\delta} {\delta \sigma}
+\frac \delta {\delta\eta}
+a \frac \delta {\delta \rho}\right)\right)\left.
{\Gamma}^\natural\right|_{\eta=\rho=0}= \nonumber\\ [3mm]
&&\left(\beta_{m_\psi}\,m_\psi\frac \partial {\partial m_\psi}
+\beta_v\,v \frac \partial {\partial v}
-\,\gamma_\psi{\cal N}_\psi
-\gamma_\Phi{\cal N}_\Phi
\right. \nonumber \\ [3mm]
&&\left.-\,\beta_v\int d^4x\,v\, \frac \delta
{\delta \sigma} 
+\delta\int d^4x\,a\,\frac \delta
{\delta \rho} \right. \nonumber \\ [3mm]
&&\left. +\,\beta_v \int d^4x\, \frac
\delta {\delta \eta} \right) \left.\Gamma^\natural
\right|_{\eta=\rho=0}
\,\,, \label{qbre1}
\end{eqnarray}
where ${\cal N}_\psi$ and ${\cal N}_\Phi$ are counting operators
given by (\ref{cop}) and (\ref{cop2}), respectively.

The latter can be rewritten in a more explicity form with the help
of the dimensional analysis identity
\begin{eqnarray}
\left({\cal D}+W_{\rm D}\right)\left.\Gamma^\natural
\right|_{\eta=\rho=0}=0
\,\,,
\label {dai}
\end{eqnarray}
where
\begin{eqnarray}
{\cal D}=
\sum\limits_{\mu=\kappa,\,v,\,m_\sigma,\,m_\pi,\,m_\psi}\mu\,
\frac \partial {\partial \mu}
\,\,,
\label {dop}
\end{eqnarray}
with $\kappa$ the mass scale at which the normalization conditions
defining the parameters of the quantum theory are taken.

This yields the Callan-Symanzik equation
\begin{eqnarray}
&&\left({\cal D}+\beta_{m_\psi}\,m_\psi\frac \partial {\partial m_\psi}
+\beta_v\,v \frac \partial {\partial v
}-\gamma_\psi{\cal N}_\psi \right. \nonumber\\ [3mm]
&&\left.-\,\gamma_\Phi{\cal N}_\Phi^{\rm hom}
\right)\left.\Gamma^\natural
\right|_{\eta=\rho=0}=
\left(-\left(1-\beta_v-\gamma_\Phi\right)\int d^4x\,v\, 
\frac \delta
{\delta \sigma} \right. \nonumber\\ [3mm] 
&&\left. -\,\left(1+\delta\right)\int d^4x\,a\,\frac \delta
{\delta \rho} -\left(1+\beta_v\right) \int d^4x\, \frac
\delta {\delta \eta} \right) \left.\Gamma^\natural
\right|_{\eta=\rho=0}
\,\,, \nonumber \\
\label{qbre2} 
\end{eqnarray}
where ${\cal N}_\Phi^{\rm hom}$ is given by (\ref{cop1}).



Ultraviolet dimensions of all fields are collected
in Table 1:

\begin{table}[tbh]
\centering
\begin{tabular}{|c||c|c|c|c|}
\hline
& $\psi$ & $\sigma$ & $\pi^a$ & $e_\mu^{m}$  \\ 
\hline\hline
$d_\Phi$ & $3/2$ & $1$ & $1$ & $0$  \\ 
 \hline 
\end{tabular}
\caption[t1]{ Ultraviolet
dimension $d_\Phi$.}
\label{table1}
\end{table}



\begin{thebibliography}{99}

\bibitem{wcz} B. Zumino, in ``{\em Effective Lagrangians and
Broken Symmetries}\,'', Lectures on Elementary Particles and
Quantum Field Theory, Brandeis 1970, \\
C.G. Callan, in ``{\em Scale Invariance and Physics
of the Light Cone}\,'', Particles Physics, Les Houches 1971, \\
W. Zimmermann, in ``{\em On the Trace Anomaly of the
Energy-Momentum Tensor}\,'', Quantum Theory of Particles and Fields,
World Scientific 1983;

\bibitem{ccj} C.G. Callan, J.S. Coleman and R. Jackiw,
{\em Ann. of Phys.} 59 (1970) 42;

\bibitem{cs} C.G. Callan, {\em Phys.Rev. D} 2 (1970) 1541, \\
K. Symanzik, {\em Comm.Math.Phys.} 18 (1970) 227;

\bibitem{sw} S. Weinberg, {\em Phys.Rev. D} 8 (1973) 3497; 

\bibitem{fmw} D.Z. Freedman, I.J. Muzinich and E.J. Weinberg,
{\em Ann. of Phys.} 87 (1974) 95;

\bibitem{cdj} J.C. Collins, A. Duncan and S.D. Joglekar,
{\em Phys.Rev. D} 16 (1977) 438;

\bibitem{lps} C. Lucchesi, O. Piguet and K. Sibold, {\em
Helv.Phys.Acta} 61 (1988) 321; \\
C. Lucchesi and G. Zoupanos, {\em Fortschr.Phys.} 45 (1997) 129;

\bibitem{bc}  A. Blasi and R. Collina, {\em Nucl.Phys. B} 345
(1990) 472;\\F. Delduc, C. Lucchesi, O. Piguet and S.P. Sorella, 
{\em Nucl.Phys. B} 346 (1990) 313; \\C. Lucchesi and O.
Piguet, {\em Nucl.Phys. B} 381 (1992) 281;

\bibitem{ms} N. Maggiore and S.P. Sorella, {\em Nucl.Phys. B}
377 (1992) 236;

\bibitem{lps1} C. Lucchesi, O. Piguet and S.P. Sorella,
{\em Nucl.Phys. B} 395 (1993) 325;

\bibitem{odhp} O.M. Del Cima, D.H.T. Franco, J.A. Helay\"el-Neto and
O. Piguet, {\em JHEP} 02 (1998) 002; {\em JHEP} 04 (1998) 010;
{\em Lett.Math.Phys.} 47 (1999) 265;

\bibitem{piguet} O. Piguet,{\em ``Supersymmetry, Ultraviolet
Finiteness and Grand Unification''}, hep-th/9606045; 

\bibitem{becchi} C. Becchi, {\em Comm.Math.Phys.} 47 (1973) 97;

\bibitem{kraus1} E. Kraus, {\em Z.Phys. C} 75 (1993) 741;

\bibitem{mpw} N. Maggiore, O. Piguet and S. Wolf, {\em Nucl.Phys. B}
476 (1996) 329;

\bibitem{bbbc} G. Bandelloni, C. Becchi, A. Blasi and
R. Collina, {\em Nucl.Phys. B} 197 (1982) 347;\\
C. Becchi, A. Blasi and
R. Collina, {\em Nucl.Phys. B} 274 (1986) 121;

\bibitem{qap}  J.H. Lowenstein, {\em Comm.Math.Phys.} 24
(1971) 1; 
{\em Phys.Rev. D} 4
(1971) 2281;\\Y.M.P. Lam, {\em Phys.Rev. D} 6
(1972) 2145; {\em Phys.Rev. D} 7 (1973) 2943;
\\T.E. Clark and J.H. Lowenstein, {\em Nucl.Phys. B} 113
(1976) 109;

\bibitem{Zimmerman}  W. Zimmermann, 1970 Brandeis Lectures, Lectures on
Elementary Particle and Quantum Field Theory, eds. S. Deser, M. Grisaru and
H. Pendleton (MIT Press Cambridge), {\em Ann.Phys.} 77
(1973) 536;

\bibitem{bert} F. Jegerlehner and B. Schroer, {\em Nucl.Phys. B}
68 (1974) 461;

\bibitem{hdafc} H.C.G. Caldas, D.H.T. Franco, A.L. Mota, F.A. Oliveira
and M.C. Nemes, {\em Nucl.Phys. A} 617 (1997) 464;

\bibitem{jackiw} R. Jackiw, {\em Phys.Rev. D} 3 (1970) 1343; 
{\em Phys.Rev. D} 3 (1970) 1356;

\bibitem{zuber} C. Itzykson and J.B. Zuber, ``{\em Quantum Field Theory}\,'',
McGraw-Hill 1980;

\bibitem{pigsor}  O. Piguet and S.P. Sorella, {\em ``Algebraic
Renormalization''}, Lecture Notes in Physics, m28, Springer-Verlag, Berlin,
Heidelberg, 1995;

\bibitem{low}  J.H. Lowenstein, 
{\em ``BPHZ Renormalization'' } in Renormalization
Theory, eds. G. Velo and A.S. Wightman (D. Reidel, Dordrecht, 1976);
{\em Comm.Math.Phys.} 47 (1976) 53;

\bibitem{iorio}  A. Iorio, L. O'Raifeartaigh, I. Sachs 
and C. Wiesendager, {\em Nucl.Phys. B} 495 (1997) 433;

\bibitem{dhac} D.H.T. Franco, H.C.G. Caldas, A.L. Mota and
M.C. Nemes, {\em Mod.Phys.Lett. A} 12 (1997) 1041;

\bibitem{bertlmann} R.A. Bertlmann, ``{\em Anomalies in Quantum
Field Theory}\,'', Oxford Science Publications 1996;

\bibitem{bonora} L. Bonora, P. Pasti and M. Tonin, {\em J.Math.Phys.}
27 (1986) 2259;

\bibitem{wezu} J. Wess and B. Zumino, {\em Phys.Lett. B} 37
(1971) 95.

\end{thebibliography}
\end{document}